\documentstyle[psfig]{mn}

\newif\ifAMStwofonts
\AMStwofontstrue

\def\spose#1{\hbox to 0pt{#1\hss}}
\def \da{\Delta\alpha/\alpha}
\def \siiv{Si{\sc \,iv}}
\def \dlam{\Delta\lambda/\lambda}

\title[Further constraints on variation of $\alpha$ from alkali doublet QSO
absorption lines]{Further constraints on variation of the fine structure
constant from alkali doublet QSO absorption lines\thanks{Data presented
herein were obtained at the W.M. Keck Observatory, which is operated as a
scientific partnership among the California Institute of Technology, the
University of California and the National Aeronautics and Space
Administration. The Observatory was made possible by the generous financial
support of the W.M. Keck Foundation.}}

\author[M. T. Murphy et al.]{M. T. Murphy$^1$\thanks{E-mail:
       mim@phys.unsw.edu.au (MTM)}, J. K. Webb$^1$, V. V. Flambaum$^1$,
       J. X. Prochaska$^2$ and\newauthor A. M. Wolfe$^3$\\
$^1$School of Physics, The University of New South Wales, UNSW Sydney NSW
       2052, Australia\\
$^2$The Observatories of the Carnegie Institute of Washington, 813 Santa
       Barbara St. Pasadena, CA 91101\\
$^3$Department of Physics and Center for Astrophysics and Space Sciences,
       University of California, San Diego,\\~C-0424, La Jolla, CA 920923}

\date{Accepted ---.
      Received ---;
      in original form ---}

\pagerange{\pageref{firstpage}--\pageref{lastpage}}
\pubyear{2001}

\begin{document}

\maketitle

\label{firstpage}

\begin{abstract}
Comparison of quasar absorption line spectra with laboratory spectra
provides a precise probe for variability of the fine structure constant,
$\alpha$, over cosmological time-scales. We constrain variation in $\alpha$
in 21 Keck/HIRES \siiv~absorption systems using the alkali doublet (AD)
method in which changes in $\alpha$ are related to changes in the doublet
spacing. The precision obtained with the AD method has been increased by a
factor of 3: $\da = (-0.5 \pm 1.3) \times 10^{-5}$. We also analyse
potential systematic errors in this result. Finally, we compare the AD
method with the many-multiplet method which has achieved an order of
magnitude greater precision and we discuss the future of the AD method.
\end{abstract}

\begin{keywords}
atomic data -- line: profiles -- techniques: spectroscopic -- quasars:
absorption lines -- ultraviolet: general
\end{keywords}

\section{Introduction}\label{sec:intro}
An experimental search for variation in the fine structure constant,
$\alpha \equiv e^2/\hbar c$, is strongly motivated by many modern
theories. In particular, theories attempting to unify gravitation with the
other fundamental interactions, such as superstring theory, often require
the existence of extra, `compactified', spatial dimensions. The scale size
of the extra dimensions is related to the values of the fundamental
constants in our observable 4--dimensional subspace. Thus, evolution of
these scale sizes with cosmology leads to time variation of the fundamental
constants, such as $\alpha$ (e.g. Forg\'{a}cs \& Horv\'{a}th 1979; Marciano
1984; Barrow 1987; Damour \& Polyakov 1994). Currently, it seems that there
is no theoretical mechanism for keeping the extra dimensions constant in
size (Li \& Gott 1998) and so such theories naturally predict variation of
$\alpha$.

Quasar (QSO) absorption line studies provide a powerful probe of such
variation. Savedoff (1956) first analysed doublet separations seen in
galaxy emission spectra to obtain constraints on variation of
$\alpha$. Absorption lines in intervening clouds along the line of sight to
QSOs are substantially narrower than intrinsic emission lines and therefore
provide a more precise probe of $\alpha$. Bahcall, Sargent \& Schmidt
(1967) first used alkali doublet (AD) spacings of gas seen in absorption
which seemed to be intrinsic to the QSO. They obtained the constraint $\da
\equiv (\alpha_z-\alpha_0)/\alpha_0 = (-2 \pm 5) \times 10^{-2}$ at a
redshift $z\approx 1.95$. Here, $\alpha_z$ and $\alpha_0$ are the values of
$\alpha$ at the absorption redshift, $z$, and in the laboratory
respectively. Since then, several authors (e.g. Cowie \& Songaila 1995;
Varshalovich, Panchuk \& Ivanchik 1996) have applied the AD method to
doublets of several species (e.g. C{\sc \,iv}, Si{\sc \,ii}, \siiv, Mg{\sc
\,ii}, Al{\sc \,iii} etc.) arising from intervening absorption clouds at
significantly lower redshift than the background QSO.

The most recent and stringent constraint using the AD method was obtained
by Varshalovich, Potekhin \& Ivanchik (2000, hereafter VPI00) using the
\siiv\,$\lambda$1393/1402 doublet. A change in $\alpha$ will lead to a
change in the doublet separation given by (VPI00, correcting a
typographical error)
\begin{equation}\label{eq:varsh}
\da = \frac{c_r}{2}\left[\frac{(\dlam)_z}{(\dlam)_0}-1\right]\,.
\end{equation}
Here, $(\dlam)_z$ and $(\dlam)_0$ are the relative doublet separations in
the absorption cloud (at redshift $z$) and in the laboratory respectively
and $c_r \approx 1$ is a constant taking into account higher order
relativistic corrections. From 16 absorption systems (towards 6 QSOs) they
obtained a mean $\da = (-4.6 \pm 4.3) \times 10^{-5}$ using a line fitting
method. The error quoted here is statistical only. VPI00 augment this with
an additional systematic error term, $\pm 1.4 \times 10^{-5}$, due to
uncertainties in the laboratory doublet separation, $(\dlam)_0$, assumed in
their analysis (see Table \ref{tab:wls}). This estimate of the potential
systematic error seems optimistic considering that the error in the
laboratory wavelength separation was previously quoted at $\delta(\dlam)_0
\sim 1{\rm \,m\AA}$ (Ivanchik, Potekhin \& Varshalovich 1999). From
equation \ref{eq:varsh}, the corresponding systematic error in $\da$ is
\begin{equation}\label{eq:daerr}
\left|\delta(\da)\right| \approx
-\frac{c_r}{2}\frac{\delta(\dlam)_0}{(\dlam)_0} \approx 5 \times 10^{-5}
\end{equation}
which is more consistent with the value reported by Ivanchik et al. (1999)
($\sim 8 \times 10^{-5}$) than by VPI00.

There are three ways in which the above constraints can be significantly
improved without the need for a much larger sample:
\begin{enumerate}
\item Improved spectral resolution: Many of the spectra used by VPI00 have
FWHM $\sim 20{\rm \,kms}^{-1}$ (e.g. Petitjean, Rauch \& Carswell 1994;
Varshalovich et al. 1996) while many absorption systems are known to
contain \siiv~lines with $b$-parameters (i.e. Doppler widths) $\sim 5{\rm
\,kms}^{-1}$ (e.g. Outram, Chaffee \& Carswell 1999). FWHM $\sim 7{\rm
\,kms}^{-1}$ is now routinely used for QSO observations with, for example,
the Keck/HIRES and the VLT/UVES. Therefore, such spectra offer the
opportunity to significantly improve constraints on $\da$.

\item Improved signal-to-noise ratio (SNR): Many of the spectra used by
VPI00 have SNR $\sim 15$ per pixel. With the advent of 8--10\,m telescopes,
significantly greater SNR can be achieved (e.g. Prochaska \& Wolfe
1996, 1997, 1999).

\item Improved laboratory wavelength measurements: Griesmann \& Kling
(2000) have increased the absolute precision of the laboratory wavelengths
of \siiv\,$\lambda$1393 and 1402 by more than two orders of magnitude. We
compare the new values with those used by VPI00 (Kelly 1987; Morton 1991,
1992) in Table 1. Note that even our conservative estimate of the
systematic error on $\da$ in equation \ref{eq:daerr} is too small. The new
measurements imply that the VPI00 result in equation \ref{eq:varsh} should
be corrected by $\approx 11.3 \times 10^{-5}$ to $\da \approx (7 \pm 4)
\times 10^{-5}$. If the error in the difference between the doublet
wavelengths is equal to the absolute uncertainty (i.e. $4 \times
10^{-5}{\rm \,\AA}$) then a precision limit of $\delta(\da) \approx 0.2
\times 10^{-5}$ can now be reached. This should be true of Griesmann \&
Kling's measurement since statistical errors in the line positions dominated their error budget (Ulf Griesmann, private communication).

\begin{table*}
\caption{Atomic data for the \siiv\,$\lambda$1393 and 1402 lines. We give
the ground and excited state configurations and compare the new laboratory
wavelength measurements with those used by VPI00. We also give the
oscillator strengths, $f$, used in our profile fitting algorithm and the
relativistic coefficients $q_1$ and $q_2$ used in equation
\ref{eq:omega}.}
\label{tab:wls}
\begin{center}
\begin{tabular}{ccccccccc}\hline
\multicolumn{1}{c}{Transition}&\multicolumn{1}{c}{Ground}&\multicolumn{1}{c}{Upper
state}&\multicolumn{1}{c}{Old $\lambda_0/{\rm \AA}\,^a$}&\multicolumn{1}{c}{New
$\lambda_0/{\rm \AA}\,^b$}&\multicolumn{1}{c}{New $\omega_0/{\rm cm}^{-1}\,^b$}&\multicolumn{1}{c}{$f^a$}&\multicolumn{1}{c}{$q_1/{\rm cm}^{-1}\,^c$}&\multicolumn{1}{c}{$q_2/{\rm cm}^{-1}\,^c$}\\\hline
\siiv\,$\lambda$1393&$2p^63s~^2$S$_{1/2}$&$2p^63p~^2$P$_{3/2}$&1393.755(6)&1393.76018(4)&71748.355(2)&0.5140&766&48\\
\siiv\,$\lambda$1402&&$2p^63p~^2$P$_{1/2}$&1402.770(6)&1402.77291(4)&71287.376(2)&0.2553&362&-8\\\hline
\end{tabular}
\end{center}
{\footnotesize $^a$Martin \& Zalubas (1983) and Kelly (1987); $^b$Griesmann \& Kling (2000); $^c$Dzuba,
Flambaum \& Webb (1999b)}
\end{table*}

\end{enumerate}

We apply the above improvements in the present work with the aim of
increasing both the accuracy and precision of $\da$ as measured using the
\siiv~AD method. The paper is organised as follows. Section
\ref{sec:results} describes our observations, analysis methods and
results. We investigate potential instrumental and astrophysical systematic
effects in Section \ref{sec:systematics}. We summarise our work in Section
\ref{sec:discussion} and compare our results with others in the
literature. We also compare the AD method with another method, the
many-multiplet (MM) method (Dzuba, Flambaum \& Webb 1999a,b; Webb et
al. 1999), which offers an order of magnitude greater precision.

\section{Data, analysis and results}\label{sec:results}

\subsection{Keck/HIRES observations}\label{sec:data}
All our QSO spectra were obtained at the Keck I 10\,m telescope on Mauna
Kea with the HIRES facility (Vogt et al. 1994) over several observing runs
from 1994 to 1997. The QSOs were generally quite faint ($m_{\rm V} \la
19.0$) so several $\sim$1--2 hour exposures were co-added for each
object. Most of the data were reduced using the HIRES data reduction
package written by T. Barlow, {\sc makee}. This package converts the
two-dimensional echelle images to fully reduced, one-dimensional,
wavelength--calibrated spectra.

Thorium--Argon (ThAr) calibration spectra were taken before and after the
QSO exposures and co-added to provide a calibration spectrum. ThAr lines
were selected and centroided to provide a wavelength solution. Some of the
spectra were reduced when {\sc makee} had no wavelength calibration
facility. In these cases, wavelength calibration was carried out using {\sc
iraf}\footnote{{\sc iraf} is distributed by the National Optical Astronomy
Observatories, which are operated by the Association of Universities for
Research in Astronomy, Inc., under cooperative agreement with the National
Science Foundation.} routines. Spectra not reduced in {\sc makee} were
fully reduced within {\sc iraf}.  1$\sigma$ error arrays were generated
assuming Poisson counting statistics. We fit continua to regions of each
spectrum containing either or both of the \siiv~doublet transitions by
fitting Legendre polynomials to $\sim 500{\rm ~kms}^{-1}$ sections. Full
details of the reduction procedures can be found in Prochaska \& Wolfe
(1996, 1997, 1999). Outram et al. (1999) have also kindly contributed their
spectrum of Q1759$+$75 taken in July 1997.

Our sample comprises 21 \siiv~absorption systems (towards 8 QSOs) over a
redshift range $z=2$--3 (mean $z=2.6$). The SNR per pixel ranges from
15--40 with most spectra having SNR $\sim 30$ and ${\rm FWHM}\la 7.5{\rm
~kms}^{-1}$ ($R = 34 000$). We provide an example absorption system in
Fig. 1.
\begin{figure}
\label{fig:q2348}
\centerline{\psfig{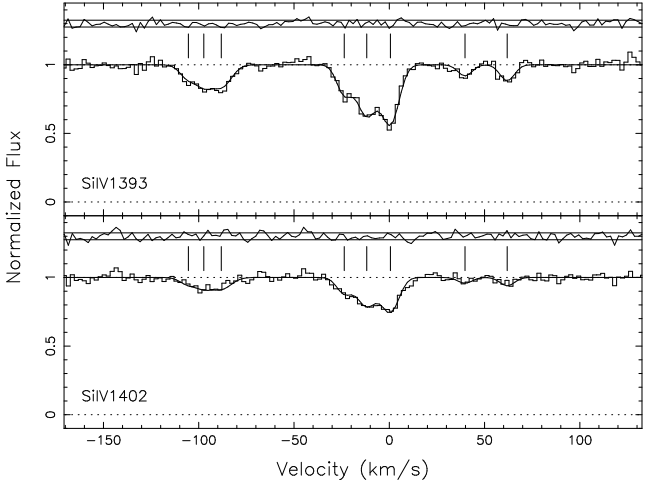}}
\caption{\siiv~absorption system at $z=2.530$ towards the QSO 2348-14. The
data have been normalized by a fit to the continuum and plotted as a
histogram. Our Voigt profile fit (solid curve) and the residuals (i.e.
$[{\rm data}] - [{\rm fit}]$), normalized to the $1\sigma$ errors
(horizontal solid lines), are also shown. The tick--marks above the
continuum indicate individual velocity components.}
\end{figure}

\subsection{Analysis}\label{sec:analysis}
Although equation \ref{eq:varsh} is a simple approach to the specific case
of an alkali doublet, a more general approach is to write down the energy
equation for an individual transition, within any multiplet and for any
species. Dzuba et al. (1999a,b) and Webb et al. (1999) suggested the
convenient formulation
\begin{equation}\label{eq:omega}
\omega_z=\omega_0+q_1x+q_2y\,,
\end{equation}
where $\omega_z$ is the wavenumber in the rest-frame of the cloud, at
redshift $z$, in which $\alpha_z/\alpha_0$ may not equal unity. $\omega_0$
is the wavenumber as measured in the laboratory and $x$ and $y$ contain the
information about $\da$:
$x\equiv\left(\frac{\alpha_z}{\alpha_0}\right)^2-1$,
$y\equiv\left(\frac{\alpha_z}{\alpha_0}\right)^4-1$. The $q_1$ and $q_2$
coefficients represent the relativistic corrections to the energy for a
particular transition. The $q_1$ coefficients are typically an order of
magnitude larger than the $q_2$ coefficients and so it is the relative
magnitudes of $q_1$ for different transitions that characterizes our
ability to constrain $\da$. This equation forms the basis of the MM method
(see Section \ref{sec:comp}). In the case of a single alkali doublet and
$\da\ll 1$, equation \ref{eq:omega} reduces to equation \ref{eq:varsh} with
\begin{equation}\label{eq:cr}
c_r \approx \frac{\delta q_1+\delta q_2}{\delta q_1 + 2\delta q_2}
\end{equation}
where $\delta q_1$ and $\delta q_2$ are the differences between the $q_1$
and $q_2$ coefficients for the doublet transitions. The values for $q_1$
and $q_2$ for \siiv\,$\lambda$1393 and 1402 have been calculated by Dzuba
et al. (1999b) and are shown in Table \ref{tab:wls}. For the \siiv~doublet,
$c_r \approx 0.9$ as used by VPI00.

Our technique is based on a simultaneous $\chi^2$ minimization analysis of
multiple component Voigt profile fits to the \siiv~absorption features in
the QSO spectra. Consider a QSO spectrum containing a single velocity
component of a specific transition. Three parameters describe the Voigt
profile fit to such a component: the column density $N$, the Doppler width
or $b$-parameter and the redshift $z$ of the absorbing gas cloud. For the
present case, we add another free parameter to the fit: $\da$.

We have used the program {\sc vpfit}\footnote{See
http://www.ast.cam.ac.uk/$^{\sim}$rfc/vpfit.html for details about
obtaining {\sc vpfit}.} (Webb 1987) to fit absorption profiles to the
spectra. We have modified {\sc vpfit} to include $\da$ as a free
parameter. Parameter errors can be calculated from the diagonal terms of
the final parameter covariance matrix (Fisher 1958). The reliability of these
errors has been confirmed using Monte Carlo simulations of a variety of
different combinations of transitions and velocity structures.

{\sc vpfit} imposes a cut-off point in parameter space such that very weak
velocity components are removed from the fit when they no longer
significantly affect the value of $\chi^2$. Conceivably, dropping even very
weak components could affect our determination of $\da$. Therefore, in such
cases we observed the trend in the values of $(\da)_i$ at each iteration
$i$ of the minimization routine to see if this trend was significantly
altered due to line dropping. If components were dropped during a fit then
we also re-ran the {\sc vpfit} algorithm, keeping the dropped components by
fixing their column density at the value just before they were dropped from
the original fit. No cases were found where the values of $\da$ from the
different runs differed significantly.

We impose several consistency checks before we accept a value of
$\da$. Firstly, the value of $\chi^2$ per degree of freedom must be
$\sim$1. Secondly, we calculated $\da$ for each absorption system using a
range of different first-guess values for $\da$ to ensure that the {\sc
vpfit} algorithm was finding global minima in the $\chi^2$ parameter space.

\subsection{Results}\label{sec:results2}
We present our results in Table \ref{tab:res}. We show our values of $\da$
for each absorption cloud together with the 1$\sigma$ error bars derived
from the {\sc vpfit} algorithm. To illustrate the distribution of $\da$
over cosmological time, we plot these results in Fig. 2 as a function of
fractional look-back time to the cloud and the absorption cloud redshift,
$z$. The weighted mean of the sample is $\da = (-0.5 \pm 1.3) \times
10^{-5}$. Our results show a 3.3-fold increase in precision over the VPI00
result. The reduced $\chi^2$ about the weighted mean value is 0.95, which
gives no evidence to suggest that we may have incorrectly estimated the
individual 1$\sigma$ errors on $\da$. We checked that the distribution of
$\da$ values shows no gross deviation from a Gaussian distribution
(although with only 21 data points this is not a rigorous check).  We also
note that the unweighted ($\da = (-0.12\pm 1.4)\times 10^{-5}$) and
weighted means are consistent with each other, again suggesting no
substantial deviation from Gaussianity.

\begin{table}
\caption{The raw results from the $\chi^2$ minimization procedure. For each
QSO sight line we identify the QSO emission redshift, $z_{\rm em}$, the
absorption cloud redshift, $z_{\rm abs}$ and the value of $\da$ for each
absorption cloud.}
\label{tab:res}
\begin{center}
\begin{tabular}{cllr}\hline
Object & $z_{\rm em}$ & $z_{\rm abs}$ & $\da /10^{-5}$\\\hline
0100$+$13 & 2.68 & 2.299     & $ -3.05\pm 7.30$\\
          &      & 2.309     & $  8.96\pm10.39$\\
0149$+$33 & 2.43 & 2.065     & $ -4.47\pm16.81$\\
          &      & 2.140     & $-12.93\pm12.69$\\
          &      & 2.204     & $ -5.89\pm 6.68$\\
0201$+$36 & 2.49 & 2.457     & $ -4.50\pm 3.57$\\
0347$-$38 & 3.24 & 2.810     & $  3.04\pm 7.43$\\
          &      & 2.899     & $-10.90\pm10.87$\\
          &      & 3.025     & $ -4.05\pm 6.78$\\
1759$+$75 & 3.05 & 2.624$^a$ & $ -5.98\pm 2.98$\\
          &      & 2.849$^a$ & $  2.56\pm 3.29$\\
          &      & 2.849     & $  2.74\pm 3.88$\\
          &      & 2.911     & $  6.00\pm 9.89$\\
          &      & 2.911$^a$ & $  0.65\pm 4.75$\\
2206$-$20 & 2.56 & 2.014     & $ -3.84\pm 6.59$\\
          &      & 2.128     & $ -3.48\pm 7.41$\\
2231$-$00 & 3.02 & 2.641     & $  0.20\pm10.65$\\
          &      & 2.986     & $  4.23\pm17.74$\\
2348$-$14 & 2.94 & 2.279     & $ 10.38\pm 6.05$\\
          &      & 2.530     & $  3.76\pm 6.53$\\
          &      & 2.775     & $ 14.04\pm 7.10$\\\hline
\end{tabular}       
\end{center}{\footnotesize
$^a$These absorbers contributed by Outram, Chaffee \& Carswell (1999)}
\end{table}

\begin{figure}
\label{fig:results}
\centerline{\psfig{file=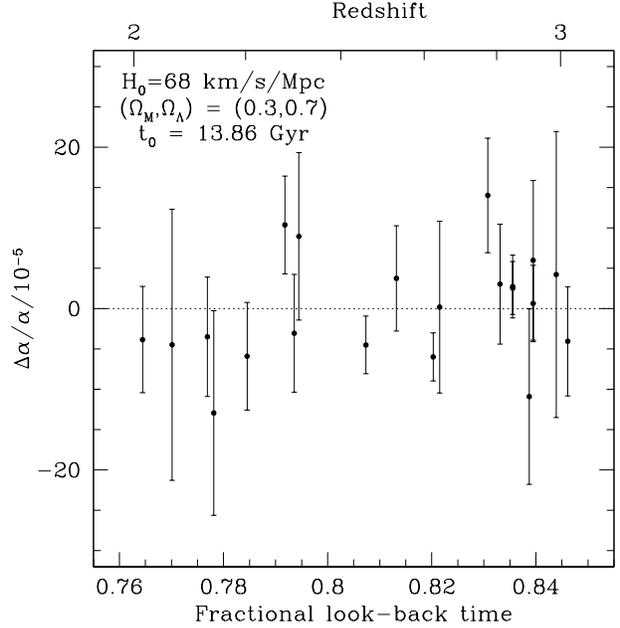,width=8.5cm}}
\caption{Our values of $\da$ for each absorption cloud plotted against
fractional look-back time (using the specified cosmological model) and
redshift. The weighted mean of the sample is $\da = (-0.5 \pm 1.3) \times
10^{-5}$.}
\end{figure}

\section{Possible systematic errors}\label{sec:systematics}
Although our result is statistically insignificant, the precision obtained
is approaching the level at which systematic errors begin to become
important. We have already discussed how the VPI00 result was severely
effected by errors in the laboratory wavelengths of \siiv\,$\lambda$1393
and 1402. We must therefore examine possible systematic errors in our
results. We have considered a range of instrumental and astrophysical
systematic errors particular to the MM method in Murphy et al. (2001b,
hereafter M01b). Below we discuss some of these in the context of the
\siiv~AD method.

\subsection{Wavelength mis-calibration}\label{sec:wavecal}
The wavelength calibration of the QSO CCD images is done by comparison with
Thorium--Argon (ThAr) lamp exposures taken before and after the QSO
frames. We consider errors in the laboratory wavelengths of the selected
ThAr lines to be negligible as their relative values are known to a similar
precision as our values of $\omega_0$ (Palmer \& Engleman Jr. 1983).
However, ThAr line mis-identifications may lead to serious mis-calibration of
the wavelength scale over large wavelength regions. Such mis-identifications
can also be applied to the rest of the spectra taken over the same series
of observations (i.e. applied to the spectra of many QSOs). Thus, {\it a
priori}, we cannot rule out the possibility that the wavelength scale has
been set improperly in this process and that a systematic shift in $\alpha$
has not been mimicked.  Such a potential effect needs careful
investigation.

We designed a direct test for wavelength mis-calibration in M01b: we treated
the ThAr emission lines in the calibration spectra with the same analysis
used to derive $\da$ from the QSO absorption lines in the object
spectra. We have applied this method to the ThAr spectra corresponding to
the absorption systems in our sample. For each absorption system we derive
values of $(\da)_{\rm ThAr}$ from corresponding sections of ThAr spectra
using the same analysis technique outlined in Section \ref{sec:analysis}
with the following modifications:
\begin{enumerate}
\item We fit Gaussian profiles to the ThAr emission lines (instead of Voigt
profiles which were fitted to the QSO absorption lines).
\item We use the literature wavenumbers of the selected ThAr lines (Palmer
\& Engleman Jr. 1983) for $\omega_0$ in equation \ref{eq:omega}. However,
we use the $q_1$ and $q_2$ coefficients of the \siiv~lines. For example,
if we select a ThAr line lying in the same region of the spectrum as the
\siiv\,$\lambda$1393 line of interest, then we assign the $q_1$ and $q_2$
coefficients for \siiv\,$\lambda$1393 to that ThAr line.
\item In M01b we noted that the $\chi^2$ for a fit to a set of ThAr lines
was $\gg 1$ (in the present case, this `set' contains two ThAr lines lying
in the same spectral regions as the corresponding \siiv~lines). This was
attributed to weak blends in the selected ThAr lines. This was also
observed in the present analysis. In this case, we cannot use the 1$\sigma$
error on $(\da)_{\rm ThAr}$ generated by {\sc vpfit}. Therefore, we select
several different sets of ThAr lines and find $(\da)_{\rm ThAr}$ for each
set. The error in $(\da)_{\rm ThAr}$ is then taken as the error in the mean
value for the different sets.
\end{enumerate}

Our results are shown in Fig. 3 where we compare them with the results from
the QSO data. It is clear that the QSO results have not been significantly
affected in any systematic way. The weighted mean value of $(\da)_{\rm
ThAr} = (-6\pm 9) \times 10^{-7}$. The scatter in the data is larger than
what we expect on the basis of the individual error bars. This could be due
to either real (low level) mis-calibrations or to the line blending
discussed above.

\begin{figure}
\label{fig:thar}
\centerline{\psfig{file=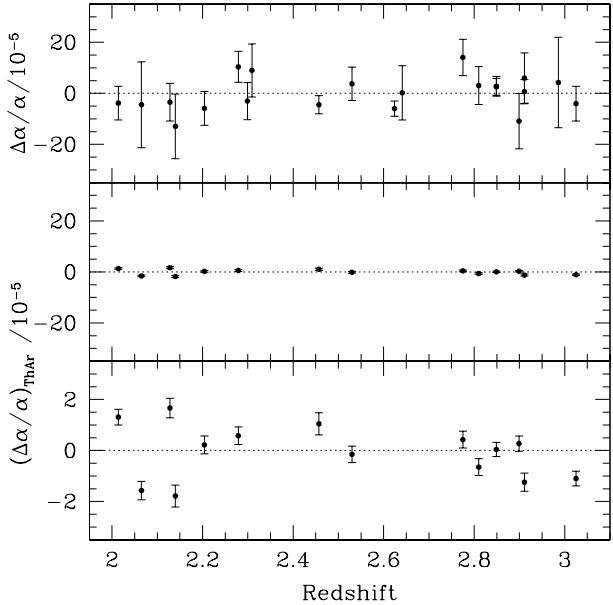,width=8.5cm}}
\caption{Comparison of QSO and ThAr results. The top panel shows the QSO
results and the middle panel shows the ThAr data on the same scale. The
weighted mean is $(-6 \pm 9)\times 10^{-7}$. The lower panel is an expanded
view of the ThAr results. Note that some ThAr data was not
available. Clearly, this does not affect our conclusions.}
\end{figure}

The results in Fig. 3 are also relevant for possible variations in
instrumental profile (IP) asymmetry along and across echelle
orders. Valenti, Butler \& Marcy (1995) have determined the HIRES IP for
several positions along a single order and they find that the IP is indeed
asymmetric and that the asymmetry varies slightly along the echelle
order. However, Fig. 3 shows that any asymmetry variations have not
systematically affected our determination of $\da$.

\begin{table*}
\caption{The isotopic structures of \siiv\,$\lambda$1393 and 1402
calculated by W. R. Johnson (private communication). We show the total
isotopic separation, $\Delta\omega_0^{\rm tot}$, and the contribution from
the specific ($\Delta\omega_0^{\rm spec}$) and volume ($\Delta\omega_0^{\rm
vol}$) isotopic shifts. The final wavenumber, $\omega_0$, is such that the
abundance-weighted mean equals the corresponding composite value (see Table
\ref{tab:wls}). The last column shows the relative abundance of each
isotope (Rosman \& Taylor 1998).}
\begin{center}
\label{tab:iso}
\begin{tabular}{lccccccr}\hline
\multicolumn{1}{c}{Transition}&\multicolumn{1}{c}{$m/{\rm amu}$}&\multicolumn{1}{c}{$\Delta\omega_0^{\rm tot}/{\rm
cm}^{-1}$}&\multicolumn{1}{c}{$\Delta\omega_0^{\rm spec}/{\rm cm}^{-1}$}&\multicolumn{1}{c}{$\Delta\omega_0^{\rm vol}/10^{-3}{\rm \,cm}^{-1}$}&\multicolumn{1}{c}{$\omega_0/{\rm
cm}^{-1}$}&\multicolumn{1}{c}{$\lambda_0/{\rm \AA}$}&\multicolumn{1}{c}{\%}\\\hline
Si{\sc \,iv}\,$\lambda$1393&28& ---  & ---  & --- &71748.344&1393.76039&92.23\\
                           &29&0.1044&0.0585& --- &71748.478&1393.75779& 4.68\\
                           &30&0.2017&0.1130&4.413&71748.545&1393.75649& 3.09\\
Si{\sc \,iv}\,$\lambda$1402&28& ---  & ---  & --- &71287.365&1402.77313&92.23\\
                           &29&0.1042&0.0583& --- &71287.499&1402.77049& 4.68\\
                           &30&0.2013&0.1126&4.418&71287.566&1402.76917& 3.09\\\hline
\end{tabular}
\end{center}
\end{table*}

\subsection{Systematic line blending with unknown species}\label{sec:blending}
There may have been weak, interloping, unresolved lines which, if the
interloping species were in the same absorption cloud, could have produced
a shift in the fitted line wavelengths of all velocity components of one or
both \siiv~transitions. We distinguish between {\it random} blends and {\it
systematic} blends. {\it Random} blends may occur if many absorption clouds
at different redshifts intersect the line of sight to a single QSO.  A {\it
systematic} blend will occur when two species are in the same cloud and
have absorption lines with similar rest-wavelengths.  Such an effect could
mimic a systematic shift in $\alpha$.

We have searched atomic line databases (Moore 1971; the Vienna Atomic Line
Database (VALD) -- Piskunov et al. 1995 and Kupka et al. 1999) for
transitions of any species which may blend with \siiv\,$\lambda$1393 or
1402. The search was restricted to transitions from the ground state with
rest wavelengths $\lambda$ such that $\left|\lambda-\lambda_0\right| \leq
0.2{\rm \,\AA}$. This is a conservative upper limit adopted from
simulations of typical blends in M01b. We did not identify any potential
blends with \siiv\,$\lambda$1402 satisfying this criteria. For
\siiv\,$\lambda$1393 we found one potential interloper: Be{\sc
\,i}\,$\lambda$1393 at $\lambda = 1393.804{\rm \,\AA}$ (VALD). This
transition has an oscillator strength of $f=2.080\times 10^{-3}$. The
Be{\sc \,i}\,$\lambda$2349 transition is nearly three orders of magnitude
stronger than this ($f=1.698$) and, to our knowledge, has never been
detected in QSO absorption spectra. This safely rules out Be{\sc
\,i}\,$\lambda$1393 as a candidate interloper.

In summary, we have found no known atomic transitions which may blend with
either of the \siiv~doublet lines. We have not considered molecular
transitions as potential interlopers but consider this possibility to be
unlikely (see M01b for discussion).

\subsection{Differential isotopic saturation}\label{sec:diffiso}
Si has three naturally occurring isotopes, $^{28}$Si, $^{29}$Si and
$^{30}$Si, with terrestrial abundances in the ratio 92.23\,:\,4.68\,:\,3.09
(Rosman \& Taylor 1998). Thus, each absorption line will be a composite of
absorption lines from all three isotopes. We are not aware of any
experimental determinations of the spectral separations between the
isotopic components. Thus, the values of $\omega_0$ in Table \ref{tab:wls}
(Griesmann \& Kling 2000) are {\it composite} wavenumbers only.  These
wavenumbers will only strictly be applicable in the optically thin regime
(linear part of the curve of growth).  As the column density increases, the
strongest isotopic component begins to saturate and the line centroid will
shift according to the natural abundances (cf. Section \ref{sec:isovar}) of
the other isotopes.

Estimates of the isotopic separations for the \siiv~doublet have been made
by W. R. Johnson (private communication) and these are summarised in Table
\ref{tab:iso}. From these separations we have estimated the absolute
isotopic wavenumbers ($\omega_0$ in Table \ref{tab:iso}) such that the
abundance-weighted mean of the isotopic wavenumbers equals the composite
values in Table \ref{tab:wls}. The error in these results is $\la 10\%$ and
comes from errors in the isotopic separations.

If we consider a \siiv~doublet with a single velocity component then, as
the $\lambda$1393 line begins to saturate, the line centroid will shift
bluewards. The effective doublet separation will continue to increase but
this increase will eventually slow and reverse as the $\lambda$1402
transition begins to saturate. It is clear that this is a potentially
important systematic effect for the \siiv~doublet. Therefore, we have used
the values of $\omega_0$ in Table \ref{tab:iso} to re-calculate $\da$ for
our sample of \siiv~absorbers. Only 5 systems clearly contain saturated
components and $\da$ in these systems displays the expected behaviour:
$\da$ becomes more negative when the isotopic structures are
included. Overall, the weighted mean for our sample becomes $\da = (-0.8
\pm 1.3) \times 10^{-5}$.

\subsection{Isotopic abundance variation}\label{sec:isovar}
Timmes \& Clayton (1996) suggest that the abundance of $^{29}$Si and
$^{30}$Si should decrease with decreasing metallicity. In general, damped
Lyman-$\alpha$ systems (DLAs), which some of our \siiv~doublets are
associated with, are thought to have metallicities $\la -1$ at high
redshift (e.g. Prochaska \& Wolfe 2000). If the isotopic abundances of Si
at redshifts $z=2$--3 differ substantially from their terrestrial values
then this may mimic a shift in $\alpha$.

The dominant component of the isotopic shift is the mass shift:
$\Delta\omega\propto\omega/m^2$ for $m$ the atomic mass. Thus, for a single
ion, the mass shift is degenerate with the redshift parameter fitted to the
velocity components of an absorption system. That is, it will have no
affect on our values of $\da$. The specific isotopic shifts for the
\siiv~doublet have been calculated by W. R. Johnson (private communication)
and are given in Table \ref{tab:iso}. We also show the volume isotopic
shift for $^{30}$\siiv~as calculated by V. A. Dzuba (private communication)
and these are clearly negligibly small. The specific shift is not
degenerate with redshift but only at a negligible level. Thus, even if the
isotopic ratios of $^{29}$Si and $^{30}$Si are much smaller in our
\siiv~absorbers, this will have a negligible affect on $\da$.

\subsection{Hyperfine structure effects}\label{sec:hyper}
Hyperfine splitting can lead to saturation effects similar to the isotopic
saturation effects discussed in Section \ref{sec:diffiso}. The transitions
of the $^{29}$Si isotope will experience hyperfine splitting since the
neutron number is odd. As with the isotopic structure, the magnitude of
hyperfine splitting has not been measured. However, V. A. Dzuba (private
communication) has estimated the splitting and finds it to be a factor of
$\sim$2 times smaller than the isotopic splitting. This may also be
estimated by comparison with the hyperfine structure of $^{25}$Mg since the
\siiv~and Mg{\sc \,ii} doublets have similar ground and excited state
wavefunctions and the magnetic moments of Mg and Si are
similar. Drullinger, Wineland \& Bergquist (1980) have measured the
hyperfine splitting of the $^{25}$Mg{\sc \,ii}\,$\lambda$2796 transition:
the hyperfine splitting is $\sim$2 times smaller than the isotopic
separations. It is therefore clear that saturation effects will be
negligible.

\subsection{Atmospheric dispersion effects}\label{sec:atmo}
As discussed in M01a, 5 of our absorption systems were observed without the
use of an image rotator (i.e. they were observed before August 1996 when a
rotator was installed on HIRES). The QSO light is dispersed across the
spectrograph slit by the atmosphere and, if the slit is not maintained
perpendicular to the horizon by an image rotator, light of different
wavelengths will enter the spectrograph at slightly different angles. In
M01b we discussed two effects resulting from this: (i) a bulk stretching of
the spectra relative to the ThAr calibration frames and (ii) a distortion
of the PSF due to truncation of the dispersed seeing disc on either side of
the slit jaws. We do not consider (ii) any further since the effect on
$\da$ found in M01b was small and since the distortion should be very
similar at both \siiv~doublet wavelengths. We consider effect (i) below.

The optical design of HIRES at the time at which our affected data was
taken was such that $\theta = \xi$ for $\theta$ the angle of the slit to
the vertical and $\xi$ the zenith angle of the object (Tom Bida, Steve
Vogt, personal communications). Consider two wavelengths, $\lambda_1$ and
$\lambda_2$ ($\lambda_2 > \lambda_1$), falling across the spectrograph
slit. If we were to measure the spectral separation between these two
wavelengths on the CCD, $\Delta\lambda'$, we would find that
\begin{equation}\label{eq:angsep}
\Delta\lambda' \approx \lambda_2-\lambda_1 +\frac{a\Delta\psi
\sin{\theta}}{\delta}\, ,
\end{equation}
where $a$ is the CCD pixel size in \AA ngstroms, $\delta$ is the projected
slit width in arc seconds per pixel (for HIRES, $\delta = 0\farcs287$ per
pixel) and $\Delta\psi$ is the angular separation (in arc seconds) of
$\lambda_1$ and $\lambda_2$ at the slit. $\Delta\psi$ is a function of the
atmospheric conditions along the line of sight to the object and can be
approximated using the refractive index of air at the observer and the
zenith distance of the object (e.g. Filippenko 1982). Note that equation
\ref{eq:angsep} is only valid if the seeing and tracking error are
zero. These two effects will reduce the measured $\Delta\lambda'$.

For the \siiv~doublet, a typical value for $\Delta\psi$ is only $\approx
0\farcs007$ (using atmospheric conditions typical of Mauna Kea, an
absorption redshift $z=2.6$ and $\psi = 30^\circ$, which is typical of our
sample, see M01b). This implies $\Delta\lambda' - (\lambda_2-\lambda_1)
\approx 3.7 \times 10^{-4}{\rm \,\AA}$ with $a=0.03{\rm \,\AA}$ per pixel
and will mimic a shift in $\alpha$, $\da \approx +0.66 \times 10^{-5}$. The
sign of this correction is such that removing the stretching implied by
equation \ref{eq:angsep} from our data will result in a more negative
$\da$. The correction is substantial for an individual system so we have
applied it to the 5 affected systems to obtain an upper limit on the
overall affect on $\da$. Our weighted mean value of $\da$ for the entire
sample becomes $(-0.6 \pm 1.3) \times 10^{-5}$.

\section{Discussion}\label{sec:discussion}

\subsection{Comparison with other results}\label{sec:comp}
We have increased the precision of $\da$ as measured with the AD method by
at least a factor of 3.3 over that of VPI00. The increase in precision (and
accuracy) is due to improved data quality (higher spectral resolution and
SNR) and greatly improved laboratory wavelength measurements (Griesmann \&
Kling 2000).

Despite this increase in precision, the MM method can provide a further
increase of up to an order of magnitude with similar quality data. The MM
method was first proposed by Dzuba, Flambaum \& Webb (1999a,b) and Webb et
al. (1999) (see also Dzuba et al. 2001). It is based on equation
\ref{eq:omega} and the use of many transitions, from many different
multiplets in different ions. The $q_1$ coefficients of transitions in
heavy ($m \sim 50$) ions are typically an order of magnitude larger than
those for light ions ($m \sim 25$). Depending on the nature of the ground
and excited state wavefunctions, the $q_1$ coefficients for some
transitions may also be of opposite sign, even for transitions of the same
ion. Therefore, the MM method has the following advantages over the AD
method:
\begin{itemize}
\item The MM method allows us to probe an effect $\sim$10 times larger
since we can compare total relativistic corrections (not just the spin
orbit coupling probed by the AD method).

\item In principle, all transitions appearing in the QSO spectra may be used
and so the number of constraints on $\da$ is increased.

\item In practice, having a larger number of transitions leads to much
better constraints on the velocity structure of the absorbing clouds.

\item Comparison of transitions with positive and negative $q_1$
coefficients minimizes the systematic effects in Section
\ref{sec:systematics}.
\end{itemize}

Webb et al. (1999) demonstrated the MM method and found tentative evidence
for a smaller $\alpha$ at redshifts $z \sim 1$: $\da = (-1.09 \pm -0.36)
\times 10^{-5}$. We have extended that work in M01a and Webb et al. 2001,
analysing 21 high redshift DLAs and re-analysing 28 lower redshift Mg/Fe
systems from the Webb et al. (1999) data set. We confirmed the Webb et
al. result and found similar results for the DLAs in Murphy et al. (2001a)
and Webb et al. (2001). For the entire sample ($0.5 < z < 3.5$), $\da =
(-0.72 \pm 0.18) \times 10^{-5}$. We analysed potential systematic errors
in M01b and found that atmospheric dispersion effects and isotopic
abundance variation may have led us to find a $\da$ that was too positive
(i.e. once these two effects are removed from our data, we find a more
negative $\da$). The results of M01a and M01b indicate possible variation
in $\alpha$. Even if these results are considered to be an upper limit on
$\da$, they are by far the most restrictive limits at high redshift. Our
present result of $\da = (-0.5 \pm 1.3) \times 10^{-5}$ is consistent with
these results. We review constraints on $\da$ at other redshifts in M01a.

\subsection{The future for the \siiv~AD method}\label{sec:future}
If the AD method were to be used to reach a similar precision as reached in
M01a, a ${\rm SNR} \ga 200$ in $\sim$50 systems would be required. In
Section \ref{sec:systematics} we found that systematic errors important for
the AD method can correspond to $\left|\da\right| \sim 0.3 \times 10^{-5}$
which is of similar magnitude to the effect of systematic errors in the MM
method (M01b). Therefore, once these systematic errors are understood and
can be reliably removed from QSO absorption data, the MM method is clearly
the preferred method for probing $\da$ at high redshift.

Despite this, a modest increase in precision could easily be achieved with
our present data if the C{\sc \,iv}\,$\lambda$1548/1550 doublet laboratory
wavelengths were known to high enough precision. Griesmann \& Kling (2000)
have increased the laboratory precision by an order of magnitude but the
statistical uncertainty in the wavelengths is still $\ga 0.04{\rm
\,cm}^{-1}$. If the laboratory precision can be increased by another order of
magnitude, the \siiv~and C{\sc \,iv} doublet can be analysed simultaneously
to obtain tighter constraints on $\da$ (i.e. the MM method could be used).

\section*{Acknowledgments}
We are very grateful to Tom Bida and Steven Vogt who provided much detailed
information about Keck/HIRES and to Ulf Griesmann and Rainer Kling for
conducting laboratory wavelength measurements specifically for the present
work. We were also motivated by the superb laboratory measurements of, and
many useful discussions with, Juliet Pickering and Anne Thorne. We thank
Bob Carswell, Fred Chaffee and Phil Outram for providing their QSO data,
Vladimir Dzuba and Walter Johnson for calculation of the isotopic and
hyperfine structure of the \siiv~doublet and Alexander Ivanchik for a
useful communication. We are grateful to the John Templeton Foundation for
supporting this work. AMW received partial NSF support from NSF grant
AST0071257.

\label{lastpage}
\end{document}